\documentclass[aip, amsmath,amssymb,
 reprint,preprintnumbers%
]{revtex4-1}

\usepackage{graphicx}
\usepackage{dcolumn}
\usepackage{bm}

\usepackage[utf8]{inputenc}
\usepackage[T1]{fontenc}
\usepackage{mathptmx}
\usepackage{etoolbox}
\usepackage{amssymb}

\usepackage{xcolor}
\usepackage[normalem]{ulem}
\usepackage{lineno}

\makeatletter
\def\@email#1#2{%
 \endgroup
 \patchcmd{\titleblock@produce}
  {\frontmatter@RRAPformat}
  {\frontmatter@RRAPformat{\produce@RRAP{*#1\href{mailto:#2}{#2}}}\frontmatter@RRAPformat}
  {}{}
}%
\makeatother
\draft 

\begin{document}


\title{Adapting reservoir computing to solve the Schrödinger equation} 



\author{L.Domingo}
\affiliation{Instituto de Ciencias Matemáticas (ICMAT); Campus de Cantoblanco UAM;
Nicolás Cabrera, 13-15; 28049 Madrid (Spain)}
\affiliation{Departamento de Química; Universidad Autónoma de Madrid;
CANTOBLANCO - 28049 Madrid (Spain)}
\affiliation{Grupo de Sistemas Complejos; Universidad Politécnica de Madrid; 28035 Madrid (Spain)}
\email[Emails: ]{laia.domingo@icmat.es; jborondo@gmail.com; f.borondo@uam.es}
\author{J. Borondo}
\affiliation{Departamento de Gestión Empresarial; Universidad Pontificia de Comillas ICADE;
  Alberto Aguilera 23; 28015 Madrid (Spain)}
\affiliation{AgrowingData; Navarro Rodrigo 2 AT; 04001 Almería (Spain)}
\author{F. Borondo}
\affiliation{Instituto de Ciencias Matemáticas (ICMAT); Campus de Cantoblanco UAM;
Nicolás Cabrera, 13-15; 28049 Madrid (Spain)}
\affiliation{Departamento de Química; Universidad Autónoma de Madrid;
CANTOBLANCO - 28049 Madrid (Spain)}


\date{\today}

\begin{abstract}
Reservoir computing is a machine learning algorithm that excels at predicting the evolution of time series, 
in particular, dynamical systems. 
Moreover, it has also shown superb performance at solving partial differential equations. 
In this work, we adapt this methodology to integrate the time-dependent Schrödinger equation, 
propagating an initial wavefunction in time.  
Since such wavefunctions are complex-valued high-dimensional arrays the reservoir computing 
formalism needs to be extended to cope with complex-valued data. 
Furthermore, we propose a multi-step learning strategy that avoids overfitting the training data. 
We illustrate the performance of our adapted reservoir computing method by application to four 
standard problems in molecular vibrational dynamics.
\end{abstract}


\maketitle 


\textbf{Reservoir computing is a machine learning  method that has proven very useful in predicting the evolution of time series, such as weather forecast, stock index or the dynamics of chaotic systems. In this work, we adapt this machine learning algorithm to predict the evolution of quantum systems, whose dynamics is determined by the Schrödinger equation. In this context, the quantum states are described using large complex-valued arrays, which makes the prediction difficult with classical reservoir computing mainly due to overfitting. For this reason, we propose a new reservoir computing-based methodology that allows to efficiently propagate quantum states in time. We illustrate the method and its performance with several quantum systems.}

\section{Introduction}
\label{Introduction}

Recurrent neural networks (RNN) \cite{RNN,LSTM} are a deep learning tool used to deal with sequential data, 
such as text or time series.
Within this context, RNNs have proven to give optimal results, due to their ability to keep the knowledge 
learnt from past data. 
However, RNNs are known to be computationally expensive during training, and such training may be complicated 
due to the problem of exploding gradients during backpropagation \cite{ConvergenceRNN}. 
To solve these problems a new trend of RNN, called \textit{Reservoir Computing (RC)}, was designed. 

Indeed, Jaeger et al \cite{ENS, ESN2} proved that as long as the reservoir fullfils certain properties, 
only training the readout layer is enough to obtain excellent performance in many tasks.
The concept of RC was proposed simultaneously as Echo State Networks (ESNs) \cite{ENS} in the field of 
machine learning, and as Liquid State Machines (LSMs) \cite{LSM} in computational neuroscience. 
Actually, reservoir computing has proven useful for chaotic time series prediction \cite{chaos1, chaos2}. 
The underlying idea is that a well-trained reservoir is able to reproduce the attractor of the originating chaotic dynamics. 

Recently, relevant advances have been made in developing RC-based algorithms. 
Multiple stacked reservoirs have been used to increase the reservoir performance 
\cite{multi-step, sub-reservoirs}. 
Also, RC has been adapted to predict time series with short training data sets \cite{ARC}. 
For predicting low dimensional time series, a RC algorithm requiring no random matrices 
and fewer hyperparameters has shown to be equivalent to standard RC \cite{next-genRC}. 

In the field of quantum chemistry, machine learning methods are proving to have advantages 
over the standard computational chemistry approaches \cite{SpecialIssue}, especially in the case
of high-dimensional systems. 
In particular, machine learning has been used to solve the time-independent Schrödinger 
equation for the electronic \cite{manyElectron, Noe} and  vibrational 
(ground state \cite{H20_NN, MLSchrodinger, DLSchrodinger} and high-energy states \cite{Domingo}) 
parts of the total wavefunction within the Born-Oppenheimer approximation.

In this work, we aim to use RC to propagate wave packets with time, that is, 
to solve the time-dependent Schrödinger equation in continuous quantum systems. 
To do so, the RC model needs to be adapted to work with wavefunctions, which are complex-valued high-dimensional matrices. 
We also propose here a new learning strategy that allows propagating wavefunctions while reducing 
the overfitting of the training data. 
Such learning strategy consists of a two-step training of the readout layer, 
where the reservoir is shown how predicting unseen data affects the evolution of the internal states. 
The learning algorithm is adapted to prevent fast error propagation during the test phase. 
A more detailed explanation of the algorithm is given in Sect.~\ref{sec:Multi-step}.

The organization of this paper is as follows. 
In Sect.~\ref{RC} we review the original formulation of RC introduced in Ref.~\onlinecite{ENS}. 
In Sect.~\ref{challenges} we present the quantum setting and the challenges that it presents 
to standard RC. 
Section~\ref{advances} describes the method we propose to overcome these challenges. 
Section~\ref{systems} describes the four quantum vibrational systems that are used for 
illustration in this work. 
The corresponding results are presented in Sect.~\ref{Results}. 
Finally, Sect.~\ref{Discussion} ends the paper by summarizing the main conclusions of the present work,
and presenting an outlook for future work..

\section{Reservoir computing}
\label{RC}

\begin{figure}
\includegraphics
   [width=0.85\columnwidth]{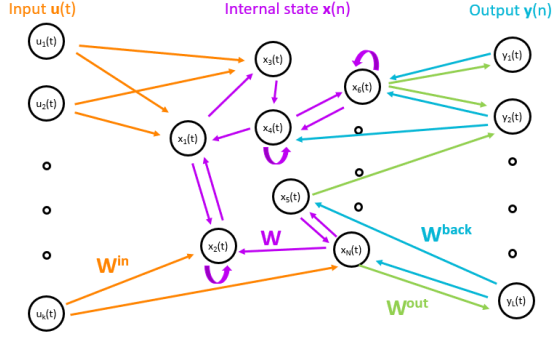}
\caption{Architecture of a reservoir computing model. Only the readout layer $W^{\text{out}}$ is learnt during training.}
\label{fig:RC}
\end{figure}

In the RC framework, the learning complexity of the algorithm is reduced to performing a linear regression.
The key point is to design a dynamical system (the internal network) that learns the input-output dynamics.
For this reason, the internal states of the network are called \textit{echo states}, 
since they can be thought of as an echo of their past \cite{ENS}. 
Such internal states are unambiguously determined by the input and output of the network. 
The structure of the network is shown in Fig.~\ref{fig:RC},%
where
\begin{itemize}
    \item $W^{\text{in}} \in M(\mathbb{R})_{N\times K}$ gives the weights from the input units to the internal states, 
    \item $W \in M(\mathbb{R})_{N\times N}$ gives the weights between the different internal states,
    \item $W^{\text{out}}\in M(\mathbb{R})_{L\times N}$ gives the wights from internal states to output units, and
    \item $W^{\text{back}}\in M(\mathbb{R})_{N\times L}$ gives the weights from the output units to the internal states,
\end{itemize}
and where
\begin{itemize}
    \item $u(t)=(u_1(t),u_2(t),\ldots,u_K(t))$ is a $K$-dimensional vector giving the input units 
    at time $t$,
    \item $x(t)=(x_1(t),x_2(t),\ldots,x_N(t)))$ is an $N$-dimensional vector giving the internal 
    states at time $t$, and
    \item $y(t)=(y_1(t),y_2(t),\ldots,y_L(t)))$ is an $L$-dimensional vector giving the output 
    units at time $t$.
\end{itemize}
Notice that the matrices $W^{\text{in}}$, $W$ and $W^{\text{back}}$ are fixed, 
so that they do not change during training. 
The only learnable parameters are the weights $W^{\text{out}}$.  
The steps to train the ESN are the following:

\begin{enumerate}
    \item Generate the reservoir matrices $(W, W^{\text{in}}, W^{\text{back}})$ randomly. 
    \item Update the internal states by teacher forcing:
    \begin{eqnarray}
    \tilde{x}(t) &&= f((W^{\text{in}} u(t)+Wx(t-1)+W^{\text{back}}y_{\text{teach}}(t-1))\nonumber\\
    x(t) &&=(1 - \alpha) x(t-1) + \alpha \tilde{x}(t)
    \label{xn_train}
    \end{eqnarray}
    where $y_{\text{teach}}$ is the output that we want out network to predict, 
    $\alpha \in (0,1]$ is the leaking rate and $f$ is the activation function. 
    Usually, $f(\cdot)=\tanh(\cdot)$, which is applied component-wise.
    \item Discard a transient of $t_\text{min}$ states to guarantee the convergence of the reservoir dynamics.
    \item Find the readout matrix $W^\text{out}$ by minimizing the mean square error:
    \begin{eqnarray}
        MSE(y, y_{\text{teach}}) = &&  \frac{1}{T - t_{\text{min}}} \sum_{t=t_{\min}}^T \Big[f^{{\text{out}}^{-1}}(y_{\text{teach}}(t)) \nonumber \\
        && - W^{\text{out}} x(t)\Big]^2 ,
    \end{eqnarray}
    where $f^{\text{out}}$ is the output activation function (usually $f^{\text{out}}= Id$).  
    Notice that this step only requires performing a linear regression.
\end{enumerate}

The steps to make the predictions after training the network are the following:
\begin{enumerate}
    \item  Given an input $u(t)$, update the state of the reservoir:
    \begin{eqnarray}
        \tilde{x}(t) &&= f[(W^{\text{in}} u(t)+Wx(t-1)+W{^\text{back}}y(t-1)]\\ \nonumber
        x(t) &&=(1 - \alpha) x(t-1) + \alpha \tilde{x}(t)
        \label{xn_test}
    \end{eqnarray}
    where $y(t-1)$ is the prediction of the output at time $t-1$.
    \item Compute the prediction $y(t)$:
    \begin{equation}
        y(t) = f^{\text{out}}\Big[W^{\text{out}} x(t)\Big].
    \end{equation}
\end{enumerate}

Let us explain now these steps in more detail. 
To ensure that the RC model works, the internal network must fulfil the \textit{echo state properties}.
Roughly speaking, this means that the internal states must have the state and input forgetting property.
That is, for $t$ large enough, $x(t)$ does not depend on $x(0), u(0)$ or $y_{\text{teach}}(0)$. 
For this reason, in step 2 of the training phase, we dismiss the initial values of the internal states,
which could be influenced by the initial parameters of the reservoir. The spectral radius $\rho(W)$ of the internal matrix $W$ also influences the performance of the method.
Larger $\rho(W)$ provide longer memory. 
However, for $\rho(W)>>1$, the reservoir will likely have no echo states. 
In fact, it is a sufficient condition that $\rho(W)<1$ to have echo states, 
even though reservoirs with $\rho(W)$ slightly larger than 1 may also give optimal results \cite{ENS}. 
Also, the matrix $W$ should be sparse and in-homogeneous so that the internal states contain 
a diverse set of trajectories. 
On the other hand, the leaking rate $\alpha$ influences the velocity of the dynamics of the output. 
A fast-changing output should be trained with greater values of $\alpha$. 
However, the RC model is also frequently used without the leaky integration, 
which is a special case obtained by setting $\alpha=1$ and thus $\tilde{x}(t) = x(t)$. 

Once the internal states have been computed, the learning phase only consists of training a 
linear model to find the mapping $W^{\text{out}}$ from the internal states to the output. 
This linear model is usually a ridge regression, which minimizes the following expression
\begin{eqnarray}
 MSE_r(y, y_{\text{teach}}) =&& \frac{1}{T - t_{\text{min}}} \sum_{t=t_{\min}}^T \Big(f^{{\text{out}}^{-1}}(y_{\text{teach}}(t)) \nonumber\\
 && - W^{\text{out}} x(t)\Big)^2 + \gamma ||W^{\text{out}}||^2 .
 \label{ridge}
 \end{eqnarray}
 This ridge regression is used to reduce overfitting during training. 
 Other methods, such as adding noise to the input during the teacher forcing phase, 
 or adding a random constant input can also be used to reduce overfitting. 
 Finally, notice that the RC model can also be used without the input layer 
 when we aim to predict a time series without explanatory variables. 
 In this case, the term $W^{\text{in}}u(t)$ is removed from Eqs.~(\ref{xn_train}) and (\ref{xn_test}).

\section{Reservoir computing challenges in quantum problems}
\label{challenges}
\subsection{The quantum setting}
In this work, we aim to solve the time-dependent Schrödinger equation given by 
\begin{equation}
    i \hbar \frac{\partial \psi(\vec{x},t)}{\partial t} = \hat{H} \psi(\vec{x},t),
    \label{eq:Schrodinger}
\end{equation}
where $\hat{H}$ is the Hamiltonian which describes the quantum system. 
The solution of Eq.~(\ref{eq:Schrodinger}), given an initial wavefunction $\psi(\vec{x},0)$, 
provides the time-evolution of the associated quantum state. 
The propagation of a quantum state with certain (mean) energy allows the calculation, 
by Fourier transformation, of the eigenenergies and eigenfunctions
of the quantum system around the same energy. 
Full details details on how to obtain the eigenenergies and eigenfunctions from $\psi(\vec{x},t)$ 
are provided in Sect.~\ref{systems}. 
Notice that this method present great advantages over the usual variational method in the case of 
excited states, since it does not require the calculation of the low lying states \cite{Domingo}.  

In this work, we aim to develop a RC model that can be trained with the short-term evolution of 
a wave packet and then use RC to predict its longer-term evolution.  
The wavefunction $\psi(\vec{x},t)$ is a complex-valued function of the spatial coordinates $\vec{x}$, 
which are usually multidimensional. 
In the RC framework, $\psi(\vec{x},t)$ is represented as a set of matrices $\{\psi(\vec{x},t)\}_t$, 
where each matrix contains the values of $\psi(\vec{x})$ at time $t$ in a grid of points spanning
$\vec{x}$. 
The use of  $\psi(\vec{x},t)$ as the target of the reservoir presents two challenges in the RC original formulation, which are described in the next subsections.

\subsection{Complex numbers}
The usual RC framework is done with real numbers. 
That is, all the inputs $u(t)$, outputs $y(t)$, internal states $x(t)$, and weight matrices $W,
W^\text{in}, W^\text{back}$ and $W^\text{out}$ are real-valued. 
However, the target time series for this quantum problem is a wavefunction $\psi(\vec{x},t)$ 
which, in general, takes complex values. 
Therefore, we need to extend update of the internal state and the learning algorithm to 
complex-valued data.

\subsection{High dimensional data}
The target data is represented as a matrix, where each entry is the value of the wavefunction 
in a discretized spatial grid. 
The size of this matrix increases exponentially 
with the dimension of the physical system, that is, the dimension of $\vec{x}$. 
In the usual RC setting, we use a large reservoir $W$, compared to the dimension of the input. 
However, in the quantum problems, the target can have a very high dimension, 
and thus it is not feasible to design a reservoir much larger than the target size. 
Moreover, the bigger the reservoir, the more data are needed to train the linear model. 
If the complexity of the method is is too large, the linear model may overfit the training data. 
Therefore, we need to adapt the RC framework to avoid this overfitting when dealing with 
high-dimensional data. 

\section{Reservoir computing advances for quantum data}
\label{advances}

This section presents the modifications done in the standard RC framework in order to adapt 
to the quantum setting.

\subsection{Complex-valued ridge regression}
We begin by presenting the extension of the RC framework to complex-valued arrays. 
The update of the internal state consists of matrix multiplications and an application of a 
non-linear function, in the following way
\begin{eqnarray}
     \tilde{x}(t) & = &f[(W^{\text{in}} u(t)+Wx(t-1)+W^{\text{back}}y_{\text{teach}}(t-1)]\nonumber\\
     x(t) & = & (1-\alpha)\;x(t-1) + \alpha \tilde{x}(t)
     \label{eq:internal_states}
\end{eqnarray}
As long as the function $f$ is defined for complex numbers, 
the previous equation holds for complex-valued arrays. 
In this work, we propose the use of two activation functions:
\begin{itemize}
    \item $f_1(x) = \tanh(x)$
    \item $f_2(x) = \tanh(\Re(x)) + i \; \tanh(\Im(x))$

\end{itemize}
Also, we set $f^\text{out} = Id$, where $Id$ is the identity function. 
Once the internal states have been calculated, the matrix $W^{\text{out}}$ is calculated 
by minimizing the MSE with $L^2$ regularization. 
This corresponds to performing a complex-valued \textit{ridge regression}
\begin{eqnarray}
    MSE_r(y, y_{\text{teach}}) = &\frac{1}{T - t_{\text{min}}} \displaystyle\sum_{t=t_{\min}}^T \Big|f_{\text{out}}^{-1}(y_{\text{teach}}(t))  \nonumber\\ & - W^{\text{out}} x(t)\Big|^2 + \alpha |W^{\text{out}}|^2,
\end{eqnarray}
where $|\cdot|$ denotes the $L^2$ norm in complex values. 
In matrix form this equation becomes
\begin{eqnarray}
    MSE_r(y, y_\text{teach}) = &(f_\text{out}^{-1}(\vec{y}_\text{teach}) - W^\text{out} X)^* \cdot (f_\text{out}^{-1}(\vec{y}_\text{teach}) - W^\text{out} X) \nonumber\\
    & + \alpha ||W^\text{out}||^2,
\end{eqnarray}
where $^*$ denotes the conjugate transpose, and $X$ is the matrix containing the internal states. 
In the case of real values, the conjugate transpose is just the transpose. 
This linear model has a closed solution
\begin{equation}
    W^\text{out} = \Big(X^* X + \alpha \mathbb{I}\Big)^{-1} \cdot \Big(X^*f_\text{out}^{-1}(\vec{y}_\text{teach})\Big),
\end{equation}
and therefore, an analytical solution for the linear model with complex-valued data can be computed. 

\subsection{Multiple-step training}
\label{sec:Multi-step}

 \begin{figure*}
\includegraphics
  [width=1.85\columnwidth]{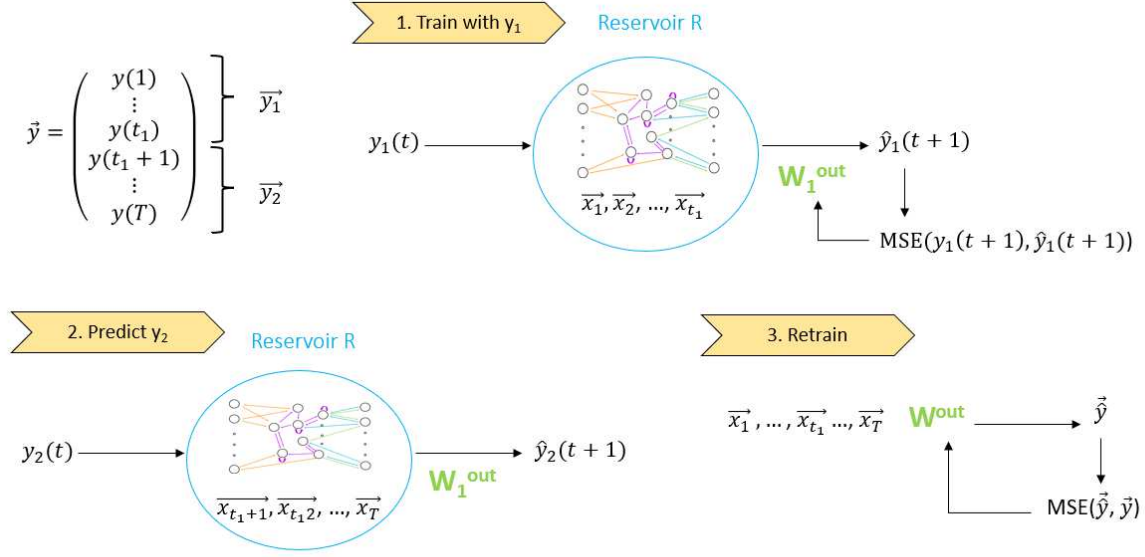}
\caption{\label{fig:MultiRC}Architecture of multi-step training of the reservoir computing model.}
\end{figure*}
The second and main challenge of dealing with quantum wavefunctions is the high-dimensionality 
of the target matrices. 
In order to capture all the underlying information of the dynamical system, the size of the reservoir $W$ should be at least of the order of magnitude of the target data. 
According to Hoeffding's theorem \cite{MLBook}, models with high complexity require a large amount 
of data to reduce the variance of the model predictions, following the (Hoeffding's) inequality 
\begin{equation}
    E_{out}\leq E_{in} + \mathcal{O}\Big(\sqrt{\frac{C}{N}}\Big),
    \label{Hoeffding}
\end{equation}
 where $E_{out}$ is the error in the test set, $E_{in}$ is the error in the training set, $C$ is a notion of complexity and $N$ is the number of data samples. 
 In our case, the size of the reservoir is similar to the number of data samples, 
 and thus the method is likely to produce overfitting. 
To reduce this effect, we propose the use of a multi-step learning training.
 The training steps in this method, depicted schematically in Fig.~\ref{fig:MultiRC}, 
 are the following:
 \begin{enumerate}
     \item Split the training input and output into two parts. 
     \begin{eqnarray}
         \vec{y} = & 
         \begin{pmatrix}
         y(1) \\
         \vdots\\
         y(t_1)\\
         y(t_1+1)\\
         \vdots \\
         y(T)
         \end{pmatrix} = 
         \begin{pmatrix}
         y_1(1)\\
         \vdots\\
         y_1(t_1)\\
         y_2(t_1 +1)\\
         \vdots\\
         y_2(T)\\
         \end{pmatrix} = 
         \begin{pmatrix}
         \vec{y_1}\\
         \vec{y_2}\\
         \end{pmatrix}, \nonumber \\
         \vec{u} = &
         \begin{pmatrix}
         u(1) \\
         \vdots\\
         u(t_1)\\
         u(t_1+1)\\
         \vdots \\
         u(T)
         \end{pmatrix} = 
         \begin{pmatrix}
         u_1(1)\\
         \vdots\\
         u_1(t_1)\\
         u_2(t_1 +1)\\
         \vdots\\
         u_2(T)\\
         \end{pmatrix} = 
         \begin{pmatrix}
         \vec{u_1}\\
         \vec{u_2}\\
         \end{pmatrix}
     \end{eqnarray}
     \item Find the internal states of the reservoir $x(1), \cdots x(t_1)$ with $\vec{y_1}$ and 
     $\vec{u_1}$, and perform a ridge regression to find $W^\text{out}$.
     \item Evolve the internal states of the reservoir $x(t_1 + 1), \ldots, x(T)$ by letting the 
     reservoir predict the values of $\vec{y_2}$.
     \item With all the internal states $x(1), \cdots, x(t_1),x(t_1+1), \ldots x(T)$ retrain the 
     linear model to correct $W^\text{out}$. 
 \end{enumerate}
When the linear model overfeeds the training data, small changes in the internal states due to 
prediction errors produce high errors in the prediction of the wavefunctions.  
Therefore, the errors tend to propagate fast until the reservoir is not able to correctly propagate 
the quantum state. 
In our proposed method, we allow the reservoir evolve after training by predicting a part of the training data. This simulates the real situation found during testing. During this predicting phase, the internal states are modified using the \textit{predicted} output instead of the \textit{exact} ones (see Eq.~\ref{xn_test}). The predicted outputs have some errors, which are transmitted to the internal states, which are the used to retrain the linear model. This learning strategy shows the reservoir how small predicting errors modify the internal states during the test phase. This reduces overfitting, decreasing the test error significantly.    

\section{Studied quantum systems}
\label{systems}

In this section we describe the quantum mechanical calculations, and we also present the quantum 
systems chosen to be studied in this work to illustrate the performance of our multi-step RC method.

\subsection{Quantum mechanical calculations}

In all cases, the training procedure is the same, and it consists of the following steps:
\begin{enumerate}
    \item Generate training and test data by numerically integrating the Schrödinger equation. 
    For this purpose, we used the widespread Fast Fourier Transform method proposed by 
    Kosloff and Kosloff \cite{kosloff}. 
    \item Train the RC and multi-step RC models. Then, compute the predicted wavefunctions
    $\psi(\vec{x},t)$ and evaluate the errors in both cases.
\end{enumerate}

From the time evolution $\psi(\vec{x},t)$ of a suitable initial quantum state $\psi_0(\vec{x})$ 
the eigenenergies and eigenfunctions of the quantum system with similar energies to that of $\psi(\vec{x},0)$ can be calculated, by subsequent Fourier transform at the quantized energies. 
The eigenenergies and eigenstates completely describe the stationary states of the quantum system. Therefore, the error of these eigenenergies gives another measure of the performance of our RC models. 
To obtain these eigenstates we have to perform two additional steps \cite{spectrum}.
\begin{enumerate}
    \setcounter{enumi}{2}
    \item Calculate the energy spectrum $I(E)$. 
    First we calculate the time-correlation function 
    \begin{equation}
        A(t) = <\psi(t)|\psi(0)> = \int_{\vec{x}} \; \psi(t)^* \psi(0) \; d\vec{x}
        \label{eq:time-corr}
    \end{equation}
Then, we obtain the energy spectrum $I(E)$ as the Fourier transform of the time-correlation function
    \begin{equation}
        I(E) = \int_t \; A(t) \; e^{i E t/ \hbar} \; dt .
    \end{equation}
    The eigenenergies $E_n$ correspond to the peaks of the energy spectrum function. 
    
    \item Calculate the eigenfunctions $\phi_n(\vec{x})$ by performing the Fourier transform of
    $\psi(\vec{x},t)$ at the eigenenergies
    \begin{equation}
        \phi_n(\vec{x}) = \int_t \; \psi(\vec{x},t) \; e^{-i E_n t/\hbar} \; dt.
    \end{equation}
    \item Evaluate the errors produced by the reservoir computing and multi-step reservoir computing models in the eigenstates and eigenenergies.  
\end{enumerate}
Now we present the different quantum systems used for this study, which include three one-dimensional quantum systems and one two-dimensional system.

\subsection{1D systems}
We aim to solve the time-dependent Schrödinger equation (\ref{eq:Schrodinger}) for 1D systems
\begin{equation}
    i \hbar \frac{\partial \psi(x,t)}{\partial t} = \hat{H} \psi(x,t)
    \label{eq:1dSchrodinger}
\end{equation}
In all cases, we set $\hbar=1$ and $m=1$ without loss of generality. 
We use 5000 time steps both for training and for test. 
The spatial domains are chosen so that they contain the wavefunctions at all times $t$. 
The systems under study are the following:
\begin{enumerate}
    \item \textbf{Harmonic oscillator}. In this case, the Hamiltonian $\hat{H}$ is
    \begin{equation}
        \hat{H}(x) = \frac{1}{2}\frac{\partial^2}{\partial x^2} +  \frac{\omega^2}{2} (x - x_0)^2 .
        \label{eq:HOsystem}
    \end{equation}
    We choose $x_0=0$ and $\omega=1$. 
    The (computationally effective) spatial domain is $[-10,10]$ covered with a grid of 200 
    equidistant points. 
    That is, $\psi(x,t)$ for a fixed $t$ is a vector of size 200. 
    The time interval between time steps is $dt=0.002$
    The initial wavefunction $\psi_0(x)$ is the minimum uncertainty Gaussian wave packet 
    \begin{equation}
        \psi(x,0) = \left(\frac{1}{\pi}\right)^{1/4} e^{-(x-x_0)^2/2} \; e^{i p_0 x},
        \label{eq:wavepacketHO}
    \end{equation}
    which is centered at $x_0=0$, and it has a width of $\displaystyle\Delta x = 1/(2\sqrt{2})$, 
    a phase/momentum $p_0 = 3.35$, and a (mean) energy $E=6.0$.

    \item \textbf{Morse potential}. The Morse Hamiltonian adequately represent the potential interaction 
    of a diatomic molecule. We write the corresponding Hamiltonian $\hat{H}$ as
    \begin{equation}
        \hat{H}(x) = \frac{1}{2}\frac{\partial^2}{\partial x^2} +  D_e\left(e^{-2a(x-x_0)} - 2 e^{-a(x-x_0)}\right),
        \label{eq:Morse}
    \end{equation}
    where $x$ is the distance between atoms, $x_0$ is the corresponding equilibrium bond distance, 
    $D_e$ is the well depth (defined relative to the dissociated atoms energy value), and $a$ is a parameter controlling the curvature of the potential at its minimum. 
    We choose $D_e=7$, $a=0.09$ and $x_0=0$. 
    The spatial domain is $[-10,25]$ with 140 points. 
    The time interval between time steps id $dt=0.007$. 
    The initial wavefunction $\psi(x,0)$ is again the minimum uncertainty packet 
    (Eq. \ref{eq:wavepacketHO}) centered at $x_0=0$, with a phase $p_0 = 2$, and an energy $E=-4.8$.

    \item \textbf{Polynomial potential} The last quantum system considered corresponds to a 
    quartic polynomial potential
    \begin{equation}
        \hat{H}(x) = \frac{1}{2}\frac{\partial^2}{\partial x^2} +  \alpha_0 + \alpha_1x + \alpha_2x^2 + \alpha_3x^3 + \alpha_4x^4,
        \label{eq:polynomial}
    \end{equation}
    where $\alpha_0 = -0.5, \alpha_1 = 0.14, \alpha_2 = 0.09, \alpha_3 = -0.01,$ and $\alpha_4 = 0.001$. 
    The spatial domain here is $[-15,25]$ with 150 points. 
    The time interval between time steps is $dt=0.007$. 
    The initial wavefunction $\psi(x,0)$ is again the minimum uncertainty packet 
    (Eq. \ref{eq:wavepacketHO}) centered at $x_0=0$, with a phase $p_0 = 2$, and an energy $E=1.8$

\end{enumerate}

\subsection{2D system}

Apart from the 1D systems we also study one 2D system, consisting of a 2D harmonic oscillator potential,
whose Hamiltonian is given by
\begin{equation}
    \hat{H}(x,y) = \frac{1}{2}\left(\frac{\partial^2}{\partial x^2}+\frac{\partial^2}{\partial y^2}\right)+  
    \frac{\omega_x^2}{2} (x - x_0)^2 + \frac{\omega_y^2}{2} (y - y_0)^2. 
\end{equation}
We choose $\omega_x^2 = 1,$ $\omega_y^2=\sqrt{2}$ and $x_0=y_0=0$. 
Notice that this choice of frequencies prevents the appearance of resonances in the dynamics and 
degeneracies in the eigenenergies.
The initial wavefunction $\psi(x,y,0)$ is the minimum uncertainty Gaussian wave packet with energy 
$E=3.2$
    \begin{equation}
        \psi_0(x,y,0) = \left(\frac{1}{\pi}\right)^{1/8} e^{-(x-x_0)^2/(4 \Delta x)^2 -(y-y_0)^2/(4 \Delta y)^2}\;
        e^{i\,p_{0}(x - y)},
        \label{eq:2DHO}
    \end{equation}
    which is centered at $x_0=y_0=0$, has widths of $\Delta x = \Delta y=0.9$, 
    and a phase $p_0 =1.75$.
    The spatial domain is $[-5,5] \times [-5,5]$ with 50 points in each dimension. 
    Therefore, $\psi(x,y,t)$ for each $t$ is a matrix of size $50 \times 50$. 
    Notice that the dimension of the data is of the order of the training data. 
    Therefore, the RC model is very likely to produce overfitting, 
    which motivates the multi-step learning we propose. 

\begin{table}
    \centering
    \begin{tabular}{lccccccc}
    \hline \hline \\
        Quantum system & $\rho(W)$ & $\alpha$ & $t_{\text{min}}$ & $\gamma$ & $N$ & $W$ density   \\ \\
         \hline  \\
         Harmonic oscillator 1D & 1.10 & 0.015 & 300 & 0.1 & 2500 & 0.015 \\ 
         Morse potential & 0.75 & 0.015 & 500 & 0.5 & 1500 & 0.015 \\
         Polynomial potential & 0.75 & 0.017 & 50 & 0.05 & 1500 & 0.008 \\
         Harmonic oscillator 2D & 1.10 & 0.06 & 300 & 0.5 & 2000 & 0.015\\
         \hline \hline
    \end{tabular}
    \caption{Reservoir computing training parameters, as defined in Sect.~\ref{RC}, for the four 
    studied quantum systems studied in this work (see Sect.~\ref{systems}).}
    \label{tab:params}
\end{table}
Table~\ref{tab:params} shows the training parameters, as defined in Sect.~\ref{RC}, 
used for each of the quantum systems considered in this work. 
The same parameter are used both for the standard RC and multi-step RC models. 
For all systems we used $f(x) = \tanh(\Re(x)) + i \; \tanh(\Im(x))$, 
and $f^\text{out} = \mathbb{I}$. 
For the multi-step RC we used 85\% of the training data for the first training step 
and 15\% for the second step.

\section{Results and discussion}
\label{Results}
In this section, we show present and discuss the results obtained with our multi-step RC method for the four models described 
in Sect.~\ref{systems}, which are compared with those rendered by the standard RC model in order to obtain an estimation 
of its performance.

\begin{table}
    \centering
    \begin{tabular}{llcc}
    \hline \hline \\
         Quantum system & Model & MSE $\psi(\vec{x},t)$ & MSE energies  \\ \\
         \hline \hline \\
         Harmonic oscillator 1D & Standard RC   &  $6 \cdot 10^{-5}$ & $ 3 \cdot 10^{-4}$\\
                                & Multi-step RC &  $2 \cdot 10^{-5}$ &  $ 1 \cdot 10^{-5}$\\
         Morse                  & Standard RC   &  $7 \cdot 10^{-4}$ & $ 4 \cdot 10^{-5}$\\
                                & Multi-step RC &  $1 \cdot 10^{-4}$ & $2 \cdot 10^{-5}$\\
         Polynomial             & Standard RC   &  $2 \cdot 10^{-4}$ & $ 3 \cdot 10^{-5}$\\
                                & Multi-step RC &  $8 \cdot 10^{-5}$ & $ 8 \cdot 10^{-6}$\\
         Harmonic oscillator 2D & Standard RC   &  $3 \cdot 10^{-4}$ & $ 4 \cdot 10^{-3}$\\
                                & Multi-step RC &  $2 \cdot 10^{-5}$ & $8 \cdot 10^{-5}$\\
         \hline \hline
    \end{tabular}
    \caption{Mean squared error (MSE) for the wavefunctions $\psi(\vec{x},t)$ and eigenenergies for the four 
    quantum systems considered in this work (see Sect. \ref{systems}) using the two studied RC models described in 
    Sects.~\ref{RC} and \ref{advances}.}
    \label{tab:mse}
\end{table}

In the first place, we show in Table~\ref{tab:mse} the mean square error (MSE) of the wavefunctions $\psi(\vec{x},t)$ 
and (mean) energies, for the different systems chosen to study. 
Let us remark that these results constitute a very strict way of gauging the performance of our method, 
since the quantum properties are evaluated in all space using $\psi^*\psi$ as the probability density, 
which means that only the regions where the wavefunction is high are meaningful. 
We see that for all cases the MSE of both the wavefunctions and energies is smaller when using the multi-step RC model. 
The largest difference in performance appears in the 2D harmonic oscillator, where the MSE is one order of magnitude smaller 
for the wavefunctions and two orders of magnitude smaller for the energies. 
Therefore, only the multi-step learning RC algorithm can correctly recover the eigenfunctions and eigenenergies of such quantum system.
Notice that the training data for the 2D harmonic oscillator consist of matrices of size $50 \times 50$ (2500 entries), 
while the 1D data are vectors of size 100 to 200. 
Therefore, it is harder to train the RC model on 2D data, since the reservoir size is  
of the same order of magnitude as the amount of training data. 
Thus, the standard RC method overfits the training data and does not generalize well to the test data. 
On the contrary, the multi-step learning strategy refits the linear readout $W^\text{out}$ by showing 
the reservoir how to adapt to unseen test data. 
This prevents having a fast error propagation during the test phase.

%
\begin{figure*}
\includegraphics
  [width=1.85\columnwidth]{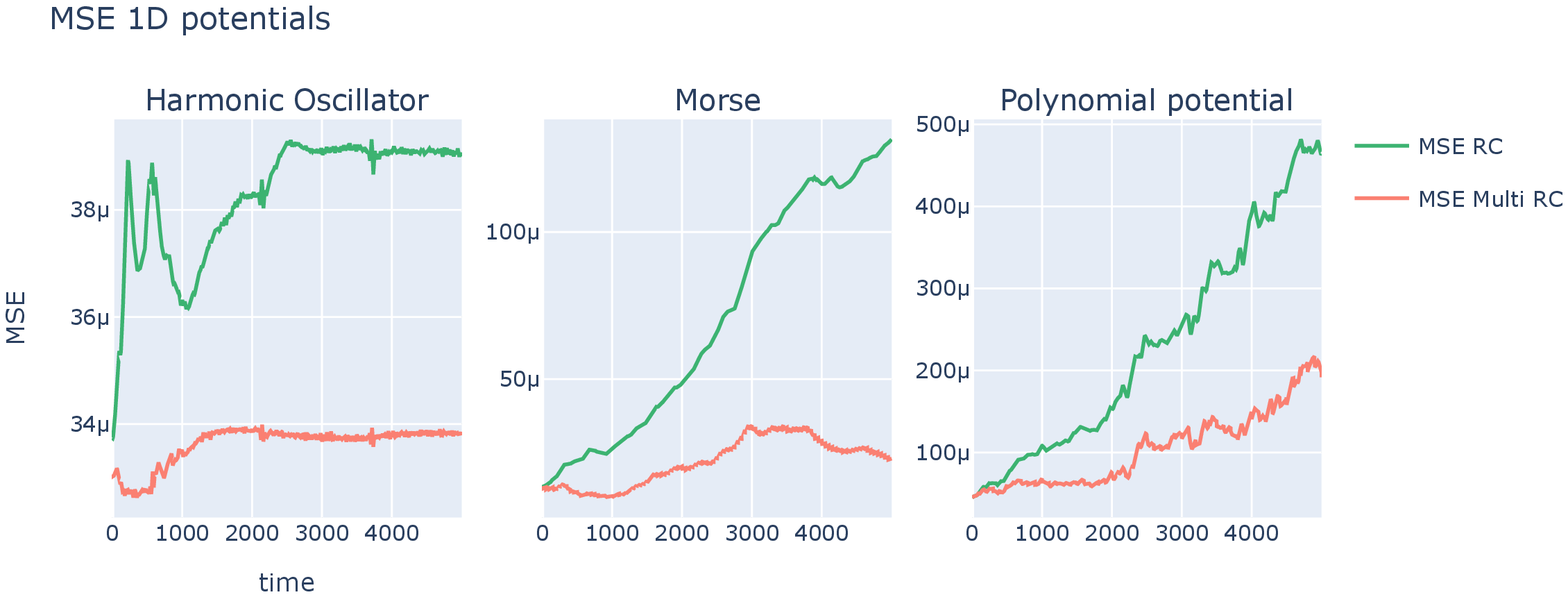}
\caption{Time evolution of the mean square error (MSE) for the harmonic, Morse, and polynomial 
oscillator time-dependent wavefunctions computed with the standard and multi-step reservoir 
computing methods.}
\label{fig:mse1D}
\end{figure*}

\begin{figure}[b!]
\includegraphics
  [width=1.00\columnwidth]{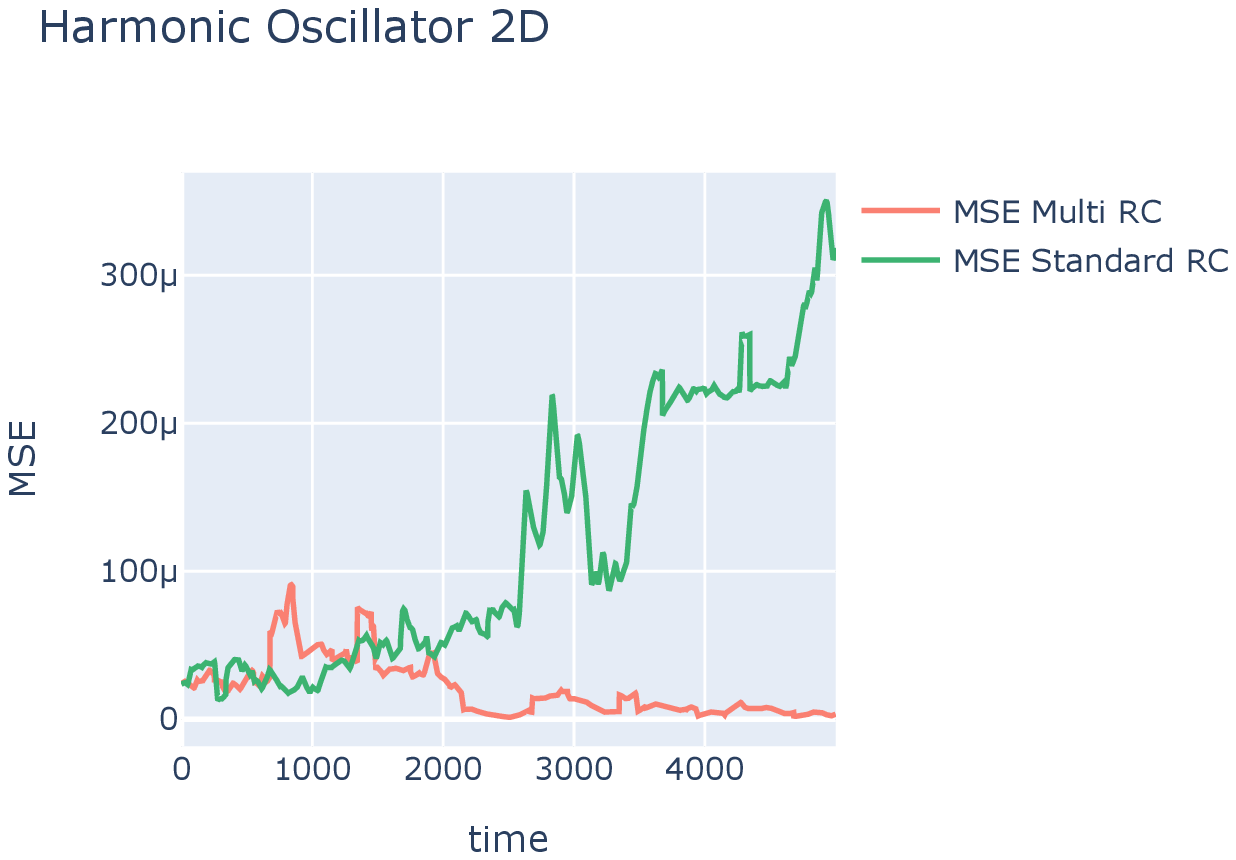}
\caption{Time evolution of the mean square error (MSE) of the 2D harmonic oscillator wavefunctions 
for the standard RC and the multi-step RC.}
\label{fig:mse2D}
\end{figure}
%
%
\begin{table}[b]
    \centering
    \begin{tabular}{ll}
    \hline \hline \\
        Quantum system & Predicted eigenenergies \\ \\
        \hline \hline \\
        Harmonic oscillator 1D & 2.5, 3.5, 4.5, 5.5, 6.5, 7.5, 8.5, 9.5 \\
        Morse & -6.8326, -6.5040, -6.1834, -5.8710, -5.5666, \\
              & -5.2704, -4.9822, -4.7022, -4.4302, -4.1664, \\
              & -3.9106, -3.6630, -3.4234, -3.1920  \\
        Polynomial potential & 0.1556, 0.6187, 1.0800, 1.5426, 2.0087, \\
                             & 2.4801, 2.9582, 3.4453 \\
        Harmonic oscillator 2D & 3.0946, 3.2838, 3.4730  \\
        \hline \hline
    \end{tabular}
    \caption{Eigenenergies obtained from the spectra calculated from the predicted wavefunctions $\psi(\vec{x},t)$ 
    computed using the multi-step RC, for all four studied quantum systems (see Sect.~\ref{systems}).}
    \label{tab:energies}
\end{table}
%
\begin{figure*}
\includegraphics[scale=0.55]{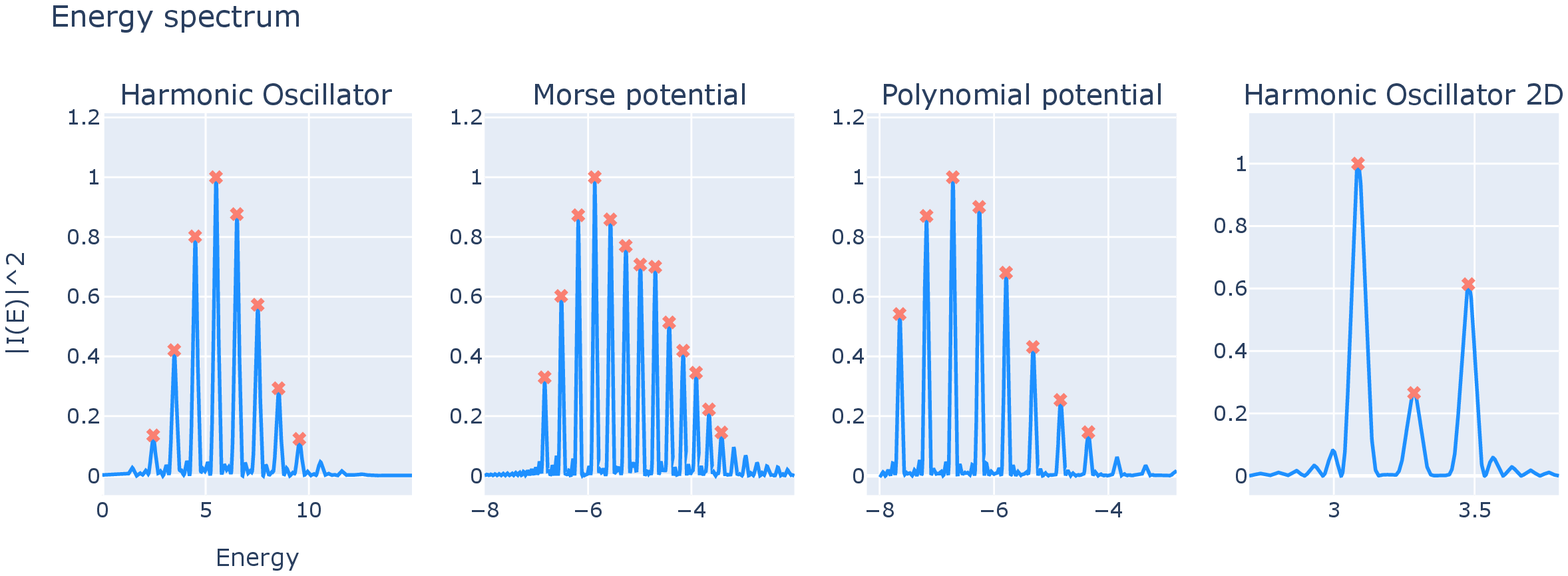}
\caption{Energy spectra obtained using the predicted wavefunctions from the multi-step RC, for all four studied quantum systems. 
The orange peaks indicate the eigenenergies of each for each of quantum system. 
Moreover, the numerical values of the eigenenergies are reported given in Table~\ref{tab:energies}.}
\label{fig:spectrum}
\end{figure*}
In Figs.~\ref{fig:mse1D} and \ref{fig:mse2D} we present the results of Table~\ref{tab:mse} in a graphic way, 
by plotting the MSE of the predicted wavefunctions $\psi(\vec{x},t)$ as a function of time. 
We see that due to error propagation the MSE tends to increase with time, except for some random fluctuations. 
For all the studied systems the MSE increases slower with time when using the multi-step RC.  
Moreover, we see that for the 1D and 2D harmonic oscillators and the Morse Hamiltonian, 
the MSE from the multi-step RC seems to stabilise, while the MSE of the standard RC keeps increasing. 
Therefore, for a fixed error tolerance, the multi-step RC method allows to predict a longer time evolution than the standard RC. 

\begin{figure*}
    \includegraphics[scale=0.7]{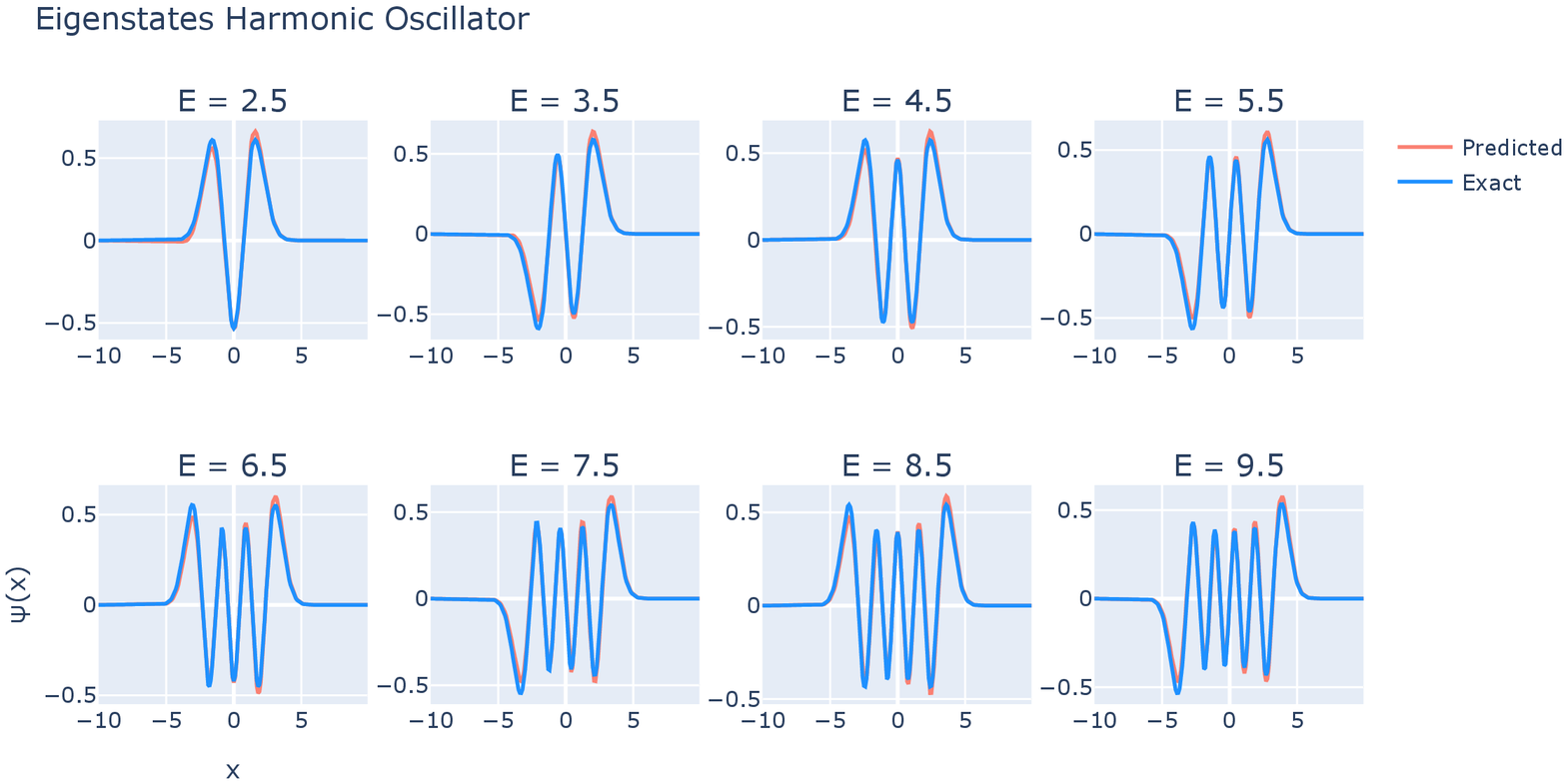}
    \caption{Exact and predicted eigenfunctions for the harmonic oscillator system (\ref{eq:HOsystem}).}
    \label{fig:wavesHO}
\end{figure*}

\begin{figure*}
    \includegraphics[scale=0.7]{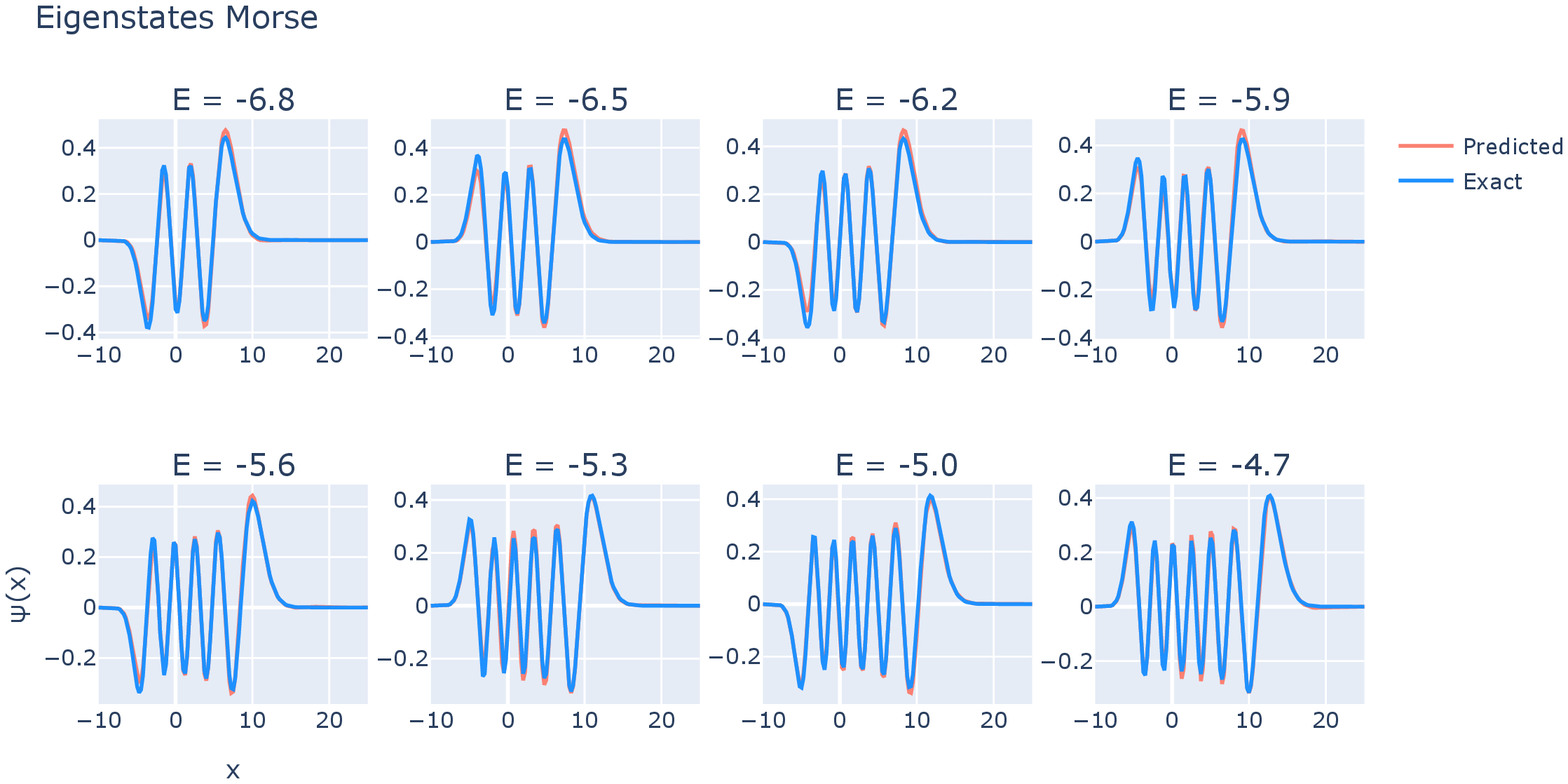}
    \caption{Same as Fig.~\ref{fig:wavesHO} for the Morse Hamiltonian system (\ref{eq:Morse}.)}
    \label{fig:wavesMorse}
\end{figure*}

\begin{figure*}
    \includegraphics[scale=0.7]{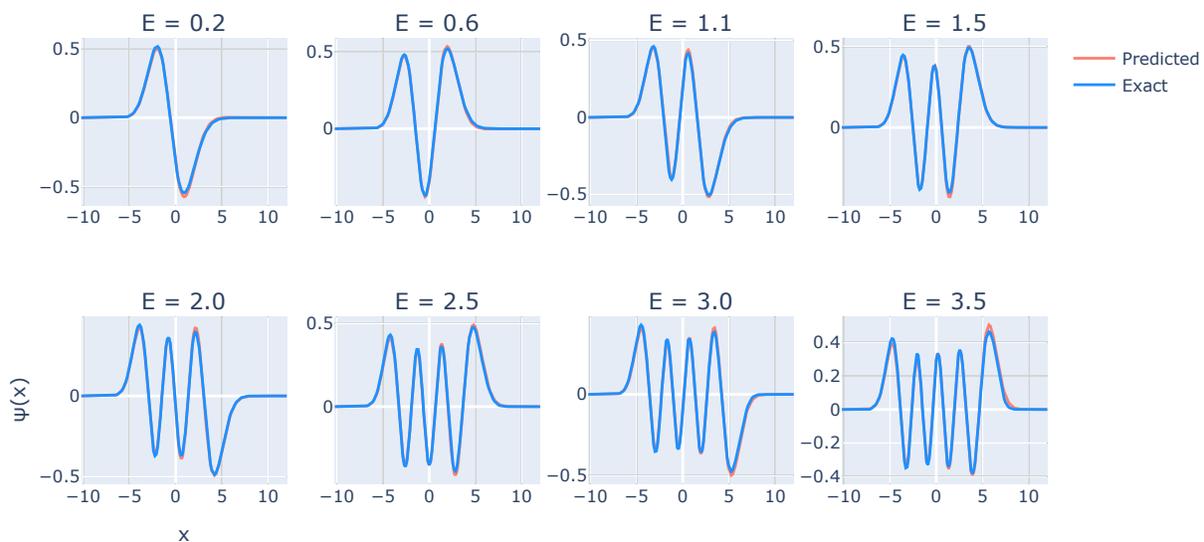}
    \caption{Same as Fig.~\ref{fig:wavesHO} for the polynomial Hamiltonian system (\ref{eq:polynomial}).}
    \label{fig:wavesPoly}
\end{figure*}

\begin{figure*}
\includegraphics[scale=0.61]{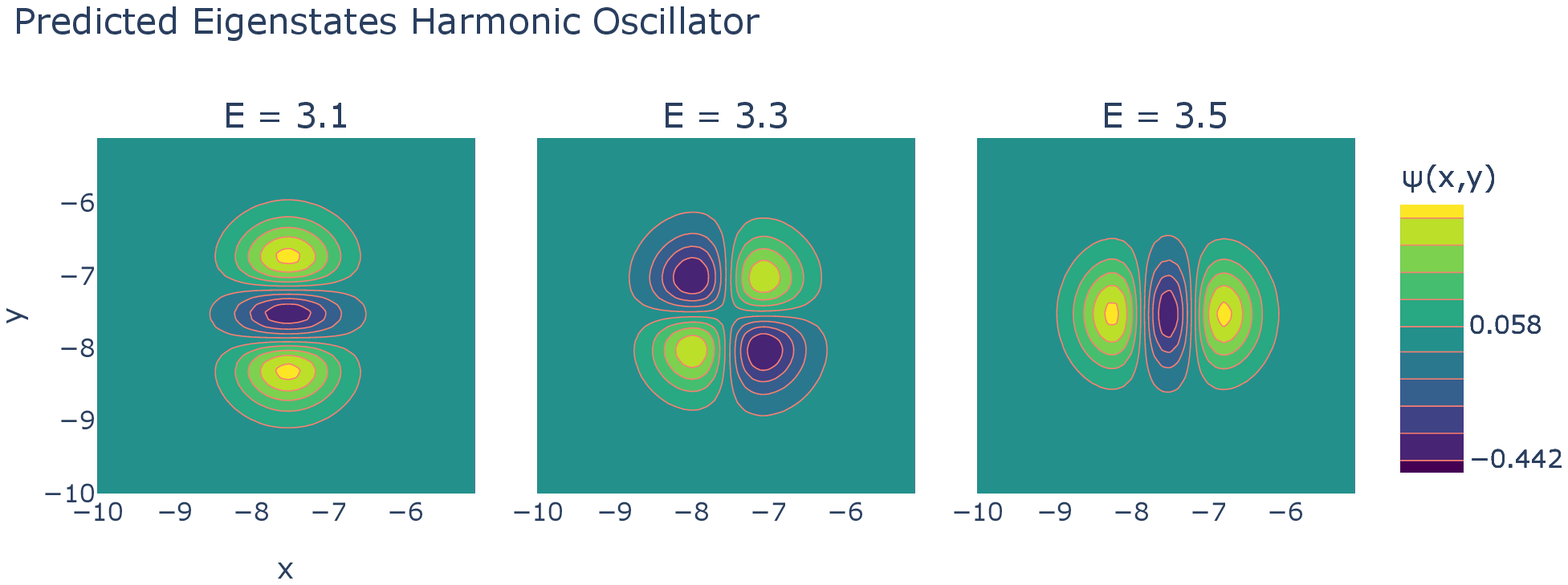}
\includegraphics[scale=0.61]{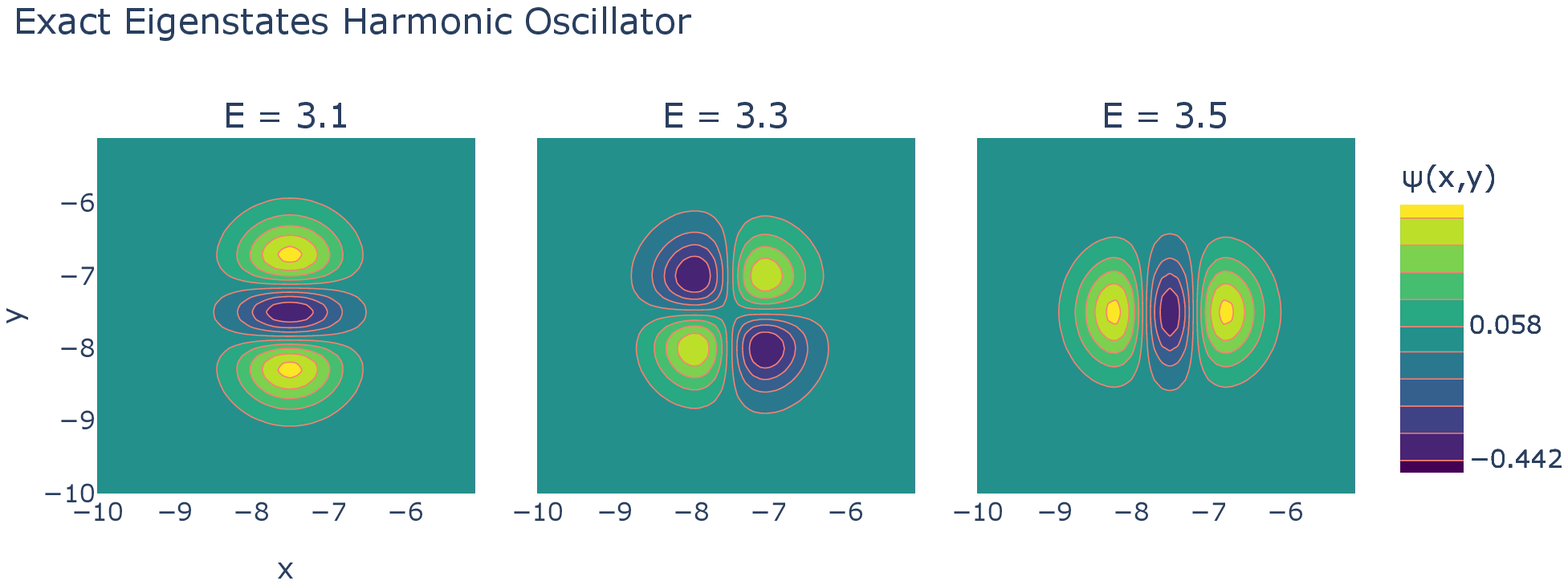}
\label{fig:wavesHO2D_exact}
\label{fig:wavesHO2D_pred}
\caption{Predicted (top row) and exact (bottom row) wavefunctions for the 2D harmonic oscillator system of
   Eq.~(\ref{eq:2DHO}).}
  \label{fig:wavesHO2D_exact}
\end{figure*}

After studying the performance of the RC models when propagating the wavefunctions in time, we can recover the eigenenergies 
and eigenfunctions around a certain energy. 
Figure~\ref{fig:spectrum} shows the spectra for the different studied systems obtained using the predicted $\psi(\vec{x},t)$ 
with the multi-step reservoir computing model. 
The peaks of the spectrum appear at the eigenenergies of each quantum system. 
These eigenenergies are close to the energy of the initial state $\psi(\vec{x},0)$ of the system. 
For example, the initial state of the 1D harmonic oscillator is a minimum uncertainty Gaussian wavepacket 
[see Eq.~(\ref{eq:wavepacketHO})] with mean energy $E=6$.  
Accordingly, the obtained eigenenergies go from $E=2.5$ to $E=9.5$, so they are centered around the initial energy $E=6$. 
Therefore, integrating the time-dependent Schrödinger equation allows recovering the eigenstates around certain energy, 
without needing to compute all the eigenstates with lower energy, as it happens in the usual variational method. 
Using the eigenenergies, we recover their associated eigenfunctions by computing the Fourier transform of the wavefunction
$\psi(\vec{x},t)$. 
Figures~\ref{fig:wavesHO}, \ref{fig:wavesMorse}, and \ref{fig:wavesPoly} show the predicted and exact eigenfunctions 
for the 1D harmonic oscillator, the Morse Hamiltonian and the polynomial potential systems, and
Fig.~\ref{fig:wavesHO2D_pred} 
shows the predicted and exact eigenfunctions for the 2D harmonic oscillator. 
The exact eigenstates are calculated analytically for the harmonic oscillator (both 1D and 2D) and the Morse potential, 
and numerically (using the variational method \cite{Domingo}) for the polynomial potential. 
We see that in all cases the predicted eigenfunctions are in very good agreement with the exact ones.  
This fact confirms that the multi-step RC method can correctly propagate a quantum wavepacket with time.
All in all, our results show that our adaptation of the RC method allows to accurately integrate the time-dependent Schrödinger equation, 
and then obtain, also with a high accuracy, the associated eigenenegies and eigenfunctions. 
This is possible thanks to the fact that our method helps prevent overfitting when working with high-dimensional 
data thus allowing the use of the adapted RC method for time propagation in quantum systems.
%
%
%

\section{Conclusions and outlook}
\label{Discussion}

Despite being based on a quite simple training framework, RC has shown remarkable performance in various benchmark tasks, 
such as time series forecasting \cite{timeSeries} and image recognition \cite{imageRecognition}.
Moreover, Pathak et al.~\cite{chaos1,chaos2} demonstrated that ESN approaches are useful for chaotic time series prediction. 

In this work, we have extended the RC framework to solve quantum problems. 
For this purpose, we propagate wavepackets with a certain energy, and then performed Fourier transform to obtain the 
eigenenergies around the initial energy and their corresponding eigenfunctions. 
The use of RC computing to integrate the time-dependent Schrödinger equation presented two main challenges. 
The first one is that the data (the quantum states $\psi(\vec{x},t)$) is complex-valued. 
To overcome this problem, we proposed to use a complex-valued activation function. 
Also, we extended the regularized linear model to complex data by extending the ridge regression to the complex domain. 
The second challenge is derived from the high-dimensionality of the input-output data. 
These data are usually represented as a matrix containing the values of the wavefunctions in a spatial grid. 
The size of the matrix increases exponentially with the dimension of the quantum system. 
Therefore, we need large reservoirs in order to propagate the quantum states, and it can easily happen that the size 
of the reservoir is similar or even larger than the number of training samples. 
In this case, the model is likely to produce overfitting, even when using a regularized model. 
When the model overfits the training data, it is unable to generalize to unseen data, and errors in the predictions propagate fast. 
To reduce the effect of this problem, we propose the use of a multi-step learning algorithm to train the RC model. 
This algorithm consists of splitting the training data into two sets. 
We first train the reservoir with the first $t_1$ time steps. 
Then, we make predictions, updating the internal states of the reservoir, with the second part of the training data. 
In this way, we teach the reservoir how the predictions affect the evolution of its internal states. 
Last, we retrain the linear readout $W^\text{out}$. 
Since the reservoir has seen how predicting modifies its internal states, small changes to the internal states due 
to prediction errors will not propagate leading to large errors in the predicted wavefunctions. 
This multi-step RC model is designed to work when the dimension of the input/output is larger or comparable to the size of the reservoir. Therefore, it could be useful to train RC models with any high-dimensional data, other than quantum wavefunctions.

As an illustration, we have applied our method to four quantum systems: three 1D systems and one 2D system. 
It is observed that the MSE of the propagated wavefunctions increases slower with time when using multi-step learning. 
This fact is critical in the 2D system, which has higher-dimensional data, this leading to more overfitting. 
In this case, the standard RC model was not able to correctly reproduce the eigenenergies of the system, 
while the multi-step learning could. 
Moreover, the multi-step learning RC also allowed to recover the eigenfunctions of all the quantum systems, 
proving that the method can correctly predict the time evolution of the wavepackets and the corresponding eigenstates. 

Once the efficiency of our multi-step RC has been proved, the present work can be extended by application to other
interesting problems. 
This future work could include, among others, the application to more complex and realistic quantum systems, 
the computation of eigenstates in a high lying energy window \cite{Revuelta2}, or the calculation of the so-called 
scarred functions \cite{Revuelta} that play a very important role in the field of quantum chaos.

\begin{acknowledgments}
The project that gave rise to these results received the support of a fellowship from "la Caixa" Foundation (ID 100010434). 
The fellowship code is LCF/BQ/DR20/11790028.
This work has also been partially supported by the Spanish Ministry of Science, Innovation and Universities, 
Gobierno de Espa\~na, under Contracts No.\ PGC2018-093854-BI00, ICMAT Severo Ochoa CEX2019-000904-S;
and by the People Programme (Marie Curie Actions) of the European Union's Horizon 2020 Research and Innovation Program 
under Grant No.~734557.
\end{acknowledgments}

\section*{Data Availability Statement}
The data that support the findings of this study are openly available in \href{https://github.com/laiadc/RC_quantum}{https://github.com/laiadc/RC\_quantum}.

\bibliography{aipsamp}

\begin{thebibliography}{26}%
\makeatletter
\providecommand \@ifxundefined [1]{%
 \@ifx{#1\undefined}
}%
\providecommand \@ifnum [1]{%
 \ifnum #1\expandafter \@firstoftwo
 \else \expandafter \@secondoftwo
 \fi
}%
\providecommand \@ifx [1]{%
 \ifx #1\expandafter \@firstoftwo
 \else \expandafter \@secondoftwo
 \fi
}%
\providecommand \natexlab [1]{#1}%
\providecommand \enquote  [1]{``#1''}%
\providecommand \bibnamefont  [1]{#1}%
\providecommand \bibfnamefont [1]{#1}%
\providecommand \citenamefont [1]{#1}%
\providecommand \href@noop [0]{\@secondoftwo}%
\providecommand \href [0]{\begingroup \@sanitize@url \@href}%
\providecommand \@href[1]{\@@startlink{#1}\@@href}%
\providecommand \@@href[1]{\endgroup#1\@@endlink}%
\providecommand \@sanitize@url [0]{\catcode `\\12\catcode `\$12\catcode
  `\&12\catcode `\#12\catcode `\^12\catcode `\_12\catcode `\%12\relax}%
\providecommand \@@startlink[1]{}%
\providecommand \@@endlink[0]{}%
\providecommand \url  [0]{\begingroup\@sanitize@url \@url }%
\providecommand \@url [1]{\endgroup\@href {#1}{\urlprefix }}%
\providecommand \urlprefix  [0]{URL }%
\providecommand \Eprint [0]{\href }%
\providecommand \doibase [0]{http://dx.doi.org/}%
\providecommand \selectlanguage [0]{\@gobble}%
\providecommand \bibinfo  [0]{\@secondoftwo}%
\providecommand \bibfield  [0]{\@secondoftwo}%
\providecommand \translation [1]{[#1]}%
\providecommand \BibitemOpen [0]{}%
\providecommand \bibitemStop [0]{}%
\providecommand \bibitemNoStop [0]{.\EOS\space}%
\providecommand \EOS [0]{\spacefactor3000\relax}%
\providecommand \BibitemShut  [1]{\csname bibitem#1\endcsname}%
\let\auto@bib@innerbib\@empty
\bibitem [{\citenamefont {Rumelhart}, \citenamefont {Hinton},\ and\
  \citenamefont {Williams}(1986)}]{RNN}%
  \BibitemOpen
  \bibfield  {author} {\bibinfo {author} {\bibfnamefont {D.~E.}\ \bibnamefont
  {Rumelhart}}, \bibinfo {author} {\bibfnamefont {G.~E.}\ \bibnamefont
  {Hinton}}, \ and\ \bibinfo {author} {\bibfnamefont {R.~J.}\ \bibnamefont
  {Williams}},\ }\bibfield  {title} {\enquote {\bibinfo {title} {Learning
  internal representations by error propagation},}\ }in\ \href@noop {} {\emph
  {\bibinfo {booktitle} {Parallel Distributed Processing: Explorations in the
  Microstructure of Cognition, {V}olume 1: {F}oundations}}}\ (\bibinfo
  {publisher} {MIT Press},\ \bibinfo {address} {Cambridge, MA},\ \bibinfo
  {year} {1986})\ pp.\ \bibinfo {pages} {318--362}\BibitemShut {NoStop}%
\bibitem [{\citenamefont {Hochreiter}\ and\ \citenamefont
  {Schmidhuber}(1997)}]{LSTM}%
  \BibitemOpen
  \bibfield  {author} {\bibinfo {author} {\bibfnamefont {S.}~\bibnamefont
  {Hochreiter}}\ and\ \bibinfo {author} {\bibfnamefont {J.}~\bibnamefont
  {Schmidhuber}},\ }\bibfield  {title} {\enquote {\bibinfo {title} {Long
  short-term memory},}\ }\href {\doibase 10.1162/neco.1997.9.8.1735} {\bibfield
   {journal} {\bibinfo  {journal} {Neural Comput.}\ }\textbf {\bibinfo {volume}
  {9}},\ \bibinfo {pages} {1735} (\bibinfo {year} {1997})}\BibitemShut
  {NoStop}%
\bibitem [{\citenamefont {Doya}(2000)}]{ConvergenceRNN}%
  \BibitemOpen
  \bibfield  {author} {\bibinfo {author} {\bibfnamefont {K.}~\bibnamefont
  {Doya}},\ }\bibfield  {title} {\enquote {\bibinfo {title} {Bifurcations in
  the learning of recurrent neural networks},}\ }\href {\doibase
  10.1109/ISCAS.1992.230622} {\bibfield  {journal} {\bibinfo  {journal} {Proc.
  IEEE Int. Symp. Circuits and Systems}\ }\textbf {\bibinfo {volume} {6}},\
  \bibinfo {pages} {2777} (\bibinfo {year} {2000})}\BibitemShut {NoStop}%
\bibitem [{\citenamefont {Jaeger}(2001)}]{ENS}%
  \BibitemOpen
  \bibfield  {author} {\bibinfo {author} {\bibfnamefont {H.}~\bibnamefont
  {Jaeger}},\ }\bibfield  {title} {\enquote {\bibinfo {title} {The "echo state"
  approach to analysing and training recurrent neural networks-with an erratum
  note},}\ }\href@noop {} {\bibfield  {journal} {\bibinfo  {journal} {German
  National Research Center for Information Technology GMD Technical Report}\
  }\textbf {\bibinfo {volume} {148}} (\bibinfo {year} {2001})}\BibitemShut
  {NoStop}%
\bibitem [{\citenamefont {Jaeger}(2007)}]{ESN2}%
  \BibitemOpen
  \bibfield  {author} {\bibinfo {author} {\bibfnamefont {H.}~\bibnamefont
  {Jaeger}},\ }\bibfield  {title} {\enquote {\bibinfo {title} {Echo state
  network},}\ }\href {\doibase 10.4249/scholarpedia.2330} {\bibfield  {journal}
  {\bibinfo  {journal} {Scholarpedia}\ }\textbf {\bibinfo {volume} {2}},\
  \bibinfo {pages} {2330} (\bibinfo {year} {2007})}\BibitemShut {NoStop}%
\bibitem [{\citenamefont {Maass}, \citenamefont {Natschläger},\ and\
  \citenamefont {Markram}(2002)}]{LSM}%
  \BibitemOpen
  \bibfield  {author} {\bibinfo {author} {\bibfnamefont {W.}~\bibnamefont
  {Maass}}, \bibinfo {author} {\bibfnamefont {T.}~\bibnamefont {Natschläger}},
  \ and\ \bibinfo {author} {\bibfnamefont {H.}~\bibnamefont {Markram}},\
  }\bibfield  {title} {\enquote {\bibinfo {title} {Real-time computing without
  stable states: a new framework for neural computation based on
  perturbations.}}\ }\href {\doibase
  https://doi.org/10.1162/089976602760407955} {\bibfield  {journal} {\bibinfo
  {journal} {Neural Comput.}\ }\textbf {\bibinfo {volume} {14}},\ \bibinfo
  {pages} {2531} (\bibinfo {year} {2002})}\BibitemShut {NoStop}%
\bibitem [{\citenamefont {Pathak}\ \emph {et~al.}(2017)\citenamefont {Pathak},
  \citenamefont {Lu}, \citenamefont {Hunt}, \citenamefont {Girvan},\ and\
  \citenamefont {Ott}}]{chaos1}%
  \BibitemOpen
  \bibfield  {author} {\bibinfo {author} {\bibfnamefont {J.}~\bibnamefont
  {Pathak}}, \bibinfo {author} {\bibfnamefont {Z.}~\bibnamefont {Lu}}, \bibinfo
  {author} {\bibfnamefont {B.}~\bibnamefont {Hunt}}, \bibinfo {author}
  {\bibfnamefont {M.}~\bibnamefont {Girvan}}, \ and\ \bibinfo {author}
  {\bibfnamefont {E.}~\bibnamefont {Ott}},\ }\bibfield  {title} {\enquote
  {\bibinfo {title} {Using machine learning to replicate chaotic attractors and
  calculate \protect{L}yapunov exponents from data},}\ }\href {\doibase
  10.1063/1.5010300} {\bibfield  {journal} {\bibinfo  {journal} {Chaos}\
  }\textbf {\bibinfo {volume} {27}},\ \bibinfo {pages} {121102} (\bibinfo
  {year} {2017})}\BibitemShut {NoStop}%
\bibitem [{\citenamefont {Pathak}\ \emph {et~al.}(2018)\citenamefont {Pathak},
  \citenamefont {Hunt}, \citenamefont {Girvan}, \citenamefont {Lu},\ and\
  \citenamefont {Ott}}]{chaos2}%
  \BibitemOpen
  \bibfield  {author} {\bibinfo {author} {\bibfnamefont {J.}~\bibnamefont
  {Pathak}}, \bibinfo {author} {\bibfnamefont {B.}~\bibnamefont {Hunt}},
  \bibinfo {author} {\bibfnamefont {M.}~\bibnamefont {Girvan}}, \bibinfo
  {author} {\bibfnamefont {Z.}~\bibnamefont {Lu}}, \ and\ \bibinfo {author}
  {\bibfnamefont {E.}~\bibnamefont {Ott}},\ }\bibfield  {title} {\enquote
  {\bibinfo {title} {Model-free prediction of large spatiotemporally chaotic
  systems from data: A reservoir computing approach},}\ }\href {\doibase
  10.1103/PhysRevLett.120.024102} {\bibfield  {journal} {\bibinfo  {journal}
  {Phys. Rev. Lett.}\ }\textbf {\bibinfo {volume} {120}},\ \bibinfo {pages}
  {024102} (\bibinfo {year} {2018})}\BibitemShut {NoStop}%
\bibitem [{\citenamefont {Akiyama}\ and\ \citenamefont
  {Tanaka}(2019)}]{multi-step}%
  \BibitemOpen
  \bibfield  {author} {\bibinfo {author} {\bibfnamefont {T.}~\bibnamefont
  {Akiyama}}\ and\ \bibinfo {author} {\bibfnamefont {G.}~\bibnamefont
  {Tanaka}},\ }\bibfield  {title} {\enquote {\bibinfo {title} {Analysis on
  characteristics of multi-step learning echo state networks for nonlinear time
  series prediction},}\ }in\ \href {\doibase 10.1109/IJCNN.2019.8851876} {\emph
  {\bibinfo {booktitle} {2019 International Joint Conference on Neural Networks
  (IJCNN)}}}\ (\bibinfo {year} {2019})\ pp.\ \bibinfo {pages}
  {1--8}\BibitemShut {NoStop}%
\bibitem [{\citenamefont {Qiao}\ \emph {et~al.}(2017)\citenamefont {Qiao},
  \citenamefont {Li}, \citenamefont {Han},\ and\ \citenamefont
  {Li}}]{sub-reservoirs}%
  \BibitemOpen
  \bibfield  {author} {\bibinfo {author} {\bibfnamefont {J.}~\bibnamefont
  {Qiao}}, \bibinfo {author} {\bibfnamefont {F.}~\bibnamefont {Li}}, \bibinfo
  {author} {\bibfnamefont {H.}~\bibnamefont {Han}}, \ and\ \bibinfo {author}
  {\bibfnamefont {W.}~\bibnamefont {Li}},\ }\bibfield  {title} {\enquote
  {\bibinfo {title} {Growing echo-state network with multiple subreservoirs},}\
  }\href {\doibase 10.1109/TNNLS.2016.2514275} {\bibfield  {journal} {\bibinfo
  {journal} {IEEE Trans. Neural Netw. Learn. Syst.}\ }\textbf {\bibinfo
  {volume} {28}},\ \bibinfo {pages} {391} (\bibinfo {year} {2017})}\BibitemShut
  {NoStop}%
\bibitem [{\citenamefont {Chen}\ \emph {et~al.}(2020)\citenamefont {Chen},
  \citenamefont {Liu}, \citenamefont {Aihara},\ and\ \citenamefont
  {Chen}}]{ARC}%
  \BibitemOpen
  \bibfield  {author} {\bibinfo {author} {\bibfnamefont {P.}~\bibnamefont
  {Chen}}, \bibinfo {author} {\bibfnamefont {R.}~\bibnamefont {Liu}}, \bibinfo
  {author} {\bibfnamefont {K.}~\bibnamefont {Aihara}}, \ and\ \bibinfo {author}
  {\bibfnamefont {L.}~\bibnamefont {Chen}},\ }\bibfield  {title} {\enquote
  {\bibinfo {title} {Autoreservoir computing for multistep ahead prediction
  based on the spatiotemporal information transformation},}\ }\href {\doibase
  10.1038/s41467-020-18381-0} {\bibfield  {journal} {\bibinfo  {journal} {Nat.
  Commun.}\ }\textbf {\bibinfo {volume} {11}},\ \bibinfo {pages} {4568}
  (\bibinfo {year} {2020})}\BibitemShut {NoStop}%
\bibitem [{\citenamefont {Gauthier}\ \emph {et~al.}(2021)\citenamefont
  {Gauthier}, \citenamefont {Bollt}, \citenamefont {Griffith},\ and\
  \citenamefont {Barbosa}}]{next-genRC}%
  \BibitemOpen
  \bibfield  {author} {\bibinfo {author} {\bibfnamefont {D.}~\bibnamefont
  {Gauthier}}, \bibinfo {author} {\bibfnamefont {E.}~\bibnamefont {Bollt}},
  \bibinfo {author} {\bibfnamefont {A.}~\bibnamefont {Griffith}}, \ and\
  \bibinfo {author} {\bibfnamefont {W.}~\bibnamefont {Barbosa}},\ }\bibfield
  {title} {\enquote {\bibinfo {title} {Next generation reservoir computing},}\
  }\href {\doibase 10.1038/s41467-021-25801-2} {\bibfield  {journal} {\bibinfo
  {journal} {Nat. Commun.}\ }\textbf {\bibinfo {volume} {12}},\ \bibinfo
  {pages} {5564} (\bibinfo {year} {2021})}\BibitemShut {NoStop}%
\bibitem [{\citenamefont {Ferguson}\ \emph {et~al.}(2020)\citenamefont
  {Ferguson}, \citenamefont {Hachmann}, \citenamefont {Miller},\ and\
  \citenamefont {Pfaendtner}}]{SpecialIssue}%
  \BibitemOpen
  \bibfield  {author} {\bibinfo {author} {\bibfnamefont {A.}~\bibnamefont
  {Ferguson}}, \bibinfo {author} {\bibfnamefont {J.}~\bibnamefont {Hachmann}},
  \bibinfo {author} {\bibfnamefont {T.}~\bibnamefont {Miller}}, \ and\ \bibinfo
  {author} {\bibfnamefont {J.}~\bibnamefont {Pfaendtner}},\ }\bibfield  {title}
  {\enquote {\bibinfo {title} {Virtual special issue on machine learning in
  physical chemistry},}\ }\href@noop {} {\bibfield  {journal} {\bibinfo
  {journal} {Journal of Physical Chemistry B}\ }\textbf {\bibinfo {volume}
  {124}},\ \bibinfo {pages} {9767} (\bibinfo {year} {2020})}\BibitemShut
  {NoStop}%
\bibitem [{\citenamefont {Pfau}\ \emph {et~al.}(2020)\citenamefont {Pfau},
  \citenamefont {Spencer}, \citenamefont {Matthews},\ and\ \citenamefont
  {Foulkes}}]{manyElectron}%
  \BibitemOpen
  \bibfield  {author} {\bibinfo {author} {\bibfnamefont {D.}~\bibnamefont
  {Pfau}}, \bibinfo {author} {\bibfnamefont {J.}~\bibnamefont {Spencer}},
  \bibinfo {author} {\bibfnamefont {A.}~\bibnamefont {Matthews}}, \ and\
  \bibinfo {author} {\bibfnamefont {W.}~\bibnamefont {Foulkes}},\ }\bibfield
  {title} {\enquote {\bibinfo {title} {Ab initio solution of the many-electron
  \protect{Schrödinger} equation with deep neural networks},}\ }\href
  {\doibase 10.1103/PhysRevResearch.2.033429} {\bibfield  {journal} {\bibinfo
  {journal} {Phys. Rev. Res.}\ }\textbf {\bibinfo {volume} {2}},\ \bibinfo
  {pages} {033429} (\bibinfo {year} {2020})}\BibitemShut {NoStop}%
\bibitem [{\citenamefont {Hermann}, \citenamefont {Sch{\"a}tzle},\ and\
  \citenamefont {No{\'e}}(2020)}]{Noe}%
  \BibitemOpen
  \bibfield  {author} {\bibinfo {author} {\bibfnamefont {J.}~\bibnamefont
  {Hermann}}, \bibinfo {author} {\bibfnamefont {Z.}~\bibnamefont
  {Sch{\"a}tzle}}, \ and\ \bibinfo {author} {\bibfnamefont {F.}~\bibnamefont
  {No{\'e}}},\ }\bibfield  {title} {\enquote {\bibinfo {title}
  {Deep-neural-network solution of the electronic \protect{Schr{\"o}dinger}
  equation},}\ }\href {\doibase https://doi.org/10.1038/s41557-020-0544-y}
  {\bibfield  {journal} {\bibinfo  {journal} {Nat. Chem.}\ }\textbf {\bibinfo
  {volume} {12}},\ \bibinfo {pages} {891} (\bibinfo {year} {2020})}\BibitemShut
  {NoStop}%
\bibitem [{\citenamefont {Manzhos}, \citenamefont {Yamashita},\ and\
  \citenamefont {Carrington}(2009)}]{H20_NN}%
  \BibitemOpen
  \bibfield  {author} {\bibinfo {author} {\bibfnamefont {S.}~\bibnamefont
  {Manzhos}}, \bibinfo {author} {\bibfnamefont {K.}~\bibnamefont {Yamashita}},
  \ and\ \bibinfo {author} {\bibfnamefont {T.}~\bibnamefont {Carrington}},\
  }\bibfield  {title} {\enquote {\bibinfo {title} {Using a neural network based
  method to solve the vibrational \protect{S}chrödinger equation for
  \protect{H$_2$O}},}\ }\href {\doibase
  https://doi.org/10.1016/j.cplett.2009.04.031} {\bibfield  {journal} {\bibinfo
   {journal} {Chem. Phys. Lett.}\ }\textbf {\bibinfo {volume} {474}},\ \bibinfo
  {pages} {217 -- 221} (\bibinfo {year} {2009})}\BibitemShut {NoStop}%
\bibitem [{\citenamefont {Pavlov}, \citenamefont {Serdyuk},\ and\ \citenamefont
  {Ustinov}(2019)}]{MLSchrodinger}%
  \BibitemOpen
  \bibfield  {author} {\bibinfo {author} {\bibfnamefont {A.}~\bibnamefont
  {Pavlov}}, \bibinfo {author} {\bibfnamefont {J.}~\bibnamefont {Serdyuk}}, \
  and\ \bibinfo {author} {\bibfnamefont {A.}~\bibnamefont {Ustinov}},\
  }\bibfield  {title} {\enquote {\bibinfo {title} {Machine learning and the
  \protect{Schrödinger} equation},}\ }\href {\doibase
  10.1088/1742-6596/1236/1/012050} {\bibfield  {journal} {\bibinfo  {journal}
  {J. Phys.: Conf. Series}\ }\textbf {\bibinfo {volume} {1236}},\ \bibinfo
  {pages} {012050} (\bibinfo {year} {2019})}\BibitemShut {NoStop}%
\bibitem [{\citenamefont {Mills}, \citenamefont {Spanner},\ and\ \citenamefont
  {Tamblyn}(2017)}]{DLSchrodinger}%
  \BibitemOpen
  \bibfield  {author} {\bibinfo {author} {\bibfnamefont {K.}~\bibnamefont
  {Mills}}, \bibinfo {author} {\bibfnamefont {M.}~\bibnamefont {Spanner}}, \
  and\ \bibinfo {author} {\bibfnamefont {I.}~\bibnamefont {Tamblyn}},\
  }\bibfield  {title} {\enquote {\bibinfo {title} {Deep learning and the
  \protect{Schr\"odinger} equation},}\ }\href {\doibase
  10.1103/PhysRevA.96.042113} {\bibfield  {journal} {\bibinfo  {journal} {Phys.
  Rev. A}\ }\textbf {\bibinfo {volume} {96}},\ \bibinfo {pages} {042113}
  (\bibinfo {year} {2017})}\BibitemShut {NoStop}%
\bibitem [{\citenamefont {Domingo}\ and\ \citenamefont
  {Borondo}(2021)}]{Domingo}%
  \BibitemOpen
  \bibfield  {author} {\bibinfo {author} {\bibfnamefont {L.}~\bibnamefont
  {Domingo}}\ and\ \bibinfo {author} {\bibfnamefont {F.}~\bibnamefont
  {Borondo}},\ }\bibfield  {title} {\enquote {\bibinfo {title} {Deep learning
  methods for the computation of vibrational wavefunctions},}\ }\href {\doibase
  https://doi.org/10.1016/j.cnsns.2021.105989} {\bibfield  {journal} {\bibinfo
  {journal} {Commun. Nonlinear Sci. Numer. Simul.}\ }\textbf {\bibinfo {volume}
  {103}},\ \bibinfo {pages} {105989} (\bibinfo {year} {2021})}\BibitemShut
  {NoStop}%
\bibitem [{\citenamefont {Mohri}, \citenamefont {Rostamizadeh},\ and\
  \citenamefont {Talwalkar}(2018)}]{MLBook}%
  \BibitemOpen
  \bibfield  {author} {\bibinfo {author} {\bibfnamefont {M.}~\bibnamefont
  {Mohri}}, \bibinfo {author} {\bibfnamefont {A.}~\bibnamefont {Rostamizadeh}},
  \ and\ \bibinfo {author} {\bibfnamefont {A.}~\bibnamefont {Talwalkar}},\
  }\href@noop {} {\emph {\bibinfo {title} {Foundations of Machine Learning}}},\
  \bibinfo {edition} {2nd}\ ed.\ (\bibinfo  {publisher} {The MIT Press},\
  \bibinfo {year} {2018})\BibitemShut {NoStop}%
\bibitem [{\citenamefont {Kosloff}\ and\ \citenamefont
  {Kosloff}(1983)}]{kosloff}%
  \BibitemOpen
  \bibfield  {author} {\bibinfo {author} {\bibfnamefont {D.}~\bibnamefont
  {Kosloff}}\ and\ \bibinfo {author} {\bibfnamefont {R.}~\bibnamefont
  {Kosloff}},\ }\bibfield  {title} {\enquote {\bibinfo {title} {A
  \protect{F}ourier method solution for the time dependent
  \protect{S}chrödinger equation as a tool in molecular dynamics},}\ }\href
  {\doibase https://doi.org/10.1016/0021-9991(83)90015-3} {\bibfield  {journal}
  {\bibinfo  {journal} {J. Comput. Phys.}\ }\textbf {\bibinfo {volume} {52}},\
  \bibinfo {pages} {35} (\bibinfo {year} {1983})}\BibitemShut {NoStop}%
\bibitem [{\citenamefont {Zdánská}\ and\ \citenamefont
  {Moiseyev}(2004)}]{spectrum}%
  \BibitemOpen
  \bibfield  {author} {\bibinfo {author} {\bibfnamefont {P.}~\bibnamefont
  {Zdánská}}\ and\ \bibinfo {author} {\bibfnamefont {N.}~\bibnamefont
  {Moiseyev}},\ }\bibfield  {title} {\enquote {\bibinfo {title} {Complex
  autocorrelation function and energy spectrum by classical trajectory
  calculations},}\ }\href {\doibase 10.1063/1.1787489} {\bibfield  {journal}
  {\bibinfo  {journal} {J. Chem. Phys.}\ }\textbf {\bibinfo {volume} {121}},\
  \bibinfo {pages} {6175} (\bibinfo {year} {2004})}\BibitemShut {NoStop}%
\bibitem [{\citenamefont {Gao}\ \emph {et~al.}(2021)\citenamefont {Gao},
  \citenamefont {Du}, \citenamefont {Duru},\ and\ \citenamefont
  {Yuen}}]{timeSeries}%
  \BibitemOpen
  \bibfield  {author} {\bibinfo {author} {\bibfnamefont {R.}~\bibnamefont
  {Gao}}, \bibinfo {author} {\bibfnamefont {L.}~\bibnamefont {Du}}, \bibinfo
  {author} {\bibfnamefont {O.}~\bibnamefont {Duru}}, \ and\ \bibinfo {author}
  {\bibfnamefont {K.~F.}\ \bibnamefont {Yuen}},\ }\bibfield  {title} {\enquote
  {\bibinfo {title} {Time series forecasting based on echo state network and
  empirical wavelet transformation},}\ }\href {\doibase
  https://doi.org/10.1016/j.asoc.2021.107111} {\bibfield  {journal} {\bibinfo
  {journal} {Appl. Soft Comput.}\ }\textbf {\bibinfo {volume} {102}},\ \bibinfo
  {pages} {107111} (\bibinfo {year} {2021})}\BibitemShut {NoStop}%
\bibitem [{\citenamefont {Jalalvand}\ \emph {et~al.}(2016)\citenamefont
  {Jalalvand}, \citenamefont {De~Neve}, \citenamefont {Van~de Walle},\ and\
  \citenamefont {Martens}}]{imageRecognition}%
  \BibitemOpen
  \bibfield  {author} {\bibinfo {author} {\bibfnamefont {A.}~\bibnamefont
  {Jalalvand}}, \bibinfo {author} {\bibfnamefont {W.}~\bibnamefont {De~Neve}},
  \bibinfo {author} {\bibfnamefont {R.}~\bibnamefont {Van~de Walle}}, \ and\
  \bibinfo {author} {\bibfnamefont {J.-P.}\ \bibnamefont {Martens}},\
  }\bibfield  {title} {\enquote {\bibinfo {title} {Towards using reservoir
  computing networks for noise-robust image recognition},}\ }in\ \href
  {\doibase 10.1109/IJCNN.2016.7727398} {\emph {\bibinfo {booktitle} {2016
  International Joint Conference on Neural Networks (IJCNN)}}}\ (\bibinfo
  {year} {2016})\ pp.\ \bibinfo {pages} {1666--1672}\BibitemShut {NoStop}%
\bibitem [{\citenamefont {Revuelta}\ \emph {et~al.}(2020)\citenamefont
  {Revuelta}, \citenamefont {Vergini}, \citenamefont {Benito},\ and\
  \citenamefont {Borondo}}]{Revuelta2}%
  \BibitemOpen
  \bibfield  {author} {\bibinfo {author} {\bibfnamefont {F.}~\bibnamefont
  {Revuelta}}, \bibinfo {author} {\bibfnamefont {E.}~\bibnamefont {Vergini}},
  \bibinfo {author} {\bibfnamefont {R.~M.}\ \bibnamefont {Benito}}, \ and\
  \bibinfo {author} {\bibfnamefont {F.}~\bibnamefont {Borondo}},\ }\bibfield
  {title} {\enquote {\bibinfo {title} {Short-periodic-orbit method for excited
  chaotic eigenfunctions},}\ }\href@noop {} {\bibfield  {journal} {\bibinfo
  {journal} {Phys. Rev. E}\ }\textbf {\bibinfo {volume} {102}} (\bibinfo {year}
  {2020})}\BibitemShut {NoStop}%
\bibitem [{\citenamefont {Revuelta}\ \emph {et~al.}(2013)\citenamefont
  {Revuelta}, \citenamefont {Benito}, \citenamefont {Borondo},\ and\
  \citenamefont {Vergini}}]{Revuelta}%
  \BibitemOpen
  \bibfield  {author} {\bibinfo {author} {\bibfnamefont {F.}~\bibnamefont
  {Revuelta}}, \bibinfo {author} {\bibfnamefont {R.}~\bibnamefont {Benito}},
  \bibinfo {author} {\bibfnamefont {F.}~\bibnamefont {Borondo}}, \ and\
  \bibinfo {author} {\bibfnamefont {E.}~\bibnamefont {Vergini}},\ }\bibfield
  {title} {\enquote {\bibinfo {title} {Using basis sets of scar functions},}\
  }\href {\doibase 10.1103/PhysRevE.87.042921} {\bibfield  {journal} {\bibinfo
  {journal} {Phys. Rev. E}\ }\textbf {\bibinfo {volume} {87}},\ \bibinfo
  {pages} {042921} (\bibinfo {year} {2013})}\BibitemShut {NoStop}%
\end{thebibliography}%

\end{document}